\documentclass[fleqn,usenatbib]{mnras}

\usepackage{newtxtext,newtxmath}
\usepackage[T1]{fontenc}
\usepackage{ae,aecompl}

\usepackage{graphicx}	% Including figure files
\usepackage{amsmath}	% Advanced maths commands
\usepackage{amssymb}	% Extra maths symbols

\newcommand\km{{\rm\thinspace km}}

\newcommand\Msun{\hbox{$\rm\thinspace M_{\odot}$}}

\newcommand\s{{\rm\thinspace s}}

\newcommand\kmps{\hbox{$\km\s^{-1}\,$}}

\title[Fast Disc Lines]{Blueshifted absorption lines from X-ray
  reflection in IRAS\,13224-3809} \author[A.C. Fabian et al] {\parbox[]{6.5in}{{ A.C.~Fabian$^1\thanks{E-mail: acf@ast.cam.ac.uk}$,
   C.S.~Reynolds$^1$, J.~Jiang$^{1,2}$,
 C.~Pinto$^3$, L.C.~Gallo$^4$, M.L.~Parker$^5$, 
A.N.~Lasenby$^{6,7}$, W.N.~Alston$^1$, D.J.K~Buisson$^8$, E.M.~Cackett$^9$,
B.~De Marco$^{10}$, J.~Garcia$^{11}$, E.~Kara$^{12}$, P.~Kosec$^1$, M.J.~Middleton$^{8}$,
 J.M.~Miller$^{13}$, G.~Miniutti$^{14}$, D.J.~Walton$^1$, 
D.R.~Wilkins$^{15}$ and A.J.~Young$^{16}$ }\\
    \footnotesize
    $^1$ Institute of Astronomy, Madingley Road, Cambridge, CB3 0HA, UK\\
$^2$ Department of Astronomy, Tsinghua University, Shuangqing Road 30, Beijing 100034, China\\
$^3$ INAF - IASF Palermo, Via U. La Malfa 153, 90146 Palermo PA, Italy\\
$^4$ Department of Astronomy and Physics, Saint Mary's University, 923
Robie Street, Halifax, NS, B3H 3C3, Canada\\
$^{5}$ European Space Agency, European Space Astronomy Centre (ESAC), E-28691 Villanueva de la Ca\~{n}ada, Madrid, Spain\\ 
$^6$ Astrophysics Group, Cavendish Laboratory, JJ Thomson Avenue,
Cambridge CB3 0HE, UK\\
$^7$ Kavli Institute for Cosmology, Institute of Astronomy, Madingley Road, Cambridge, CB3 0HA, UK\\
$^{8}$ Department of Physics and Astronomy, University of Southampton,
Highfield, Southampton SO17 1BJ, UK\\
$^9$ Department of Physics and Astronomy, Wayne State University, 666 W, Hancock Street, Detroit, MI 48201, USA\\
$^{10}$ Nicolaus Copernicus Astronomical Center, Polish Academy of Sciences, Bartycka 18, PL-00-716 Warsaw, Poland\\
$^{11}$ Cahill Center for Astronomy and Astrophysics, California Institute of Technology, Pasadena, CA 91125, USA\\
$^{12}$ MIT Kavli Institute for Astrophysics and Space Research, Cambridge, MA 02139, USA\\
$^{13}$ University of Michigan, Department of Astronomy, 1085 S. University, Ann Arbor, MI 48109, USA\\
$^{14}$ Centro de Astrobiolog\'{i}a (INTA--CSIC), Dep. de Astrof\'{i}sica, ESAC campus, 28692 Villanueva de la Ca\~nada, Spain\\
$^{15}$  Kavli Institute of Particle Astrophysics and Cosmology,
Stanford University, 452 Lomita Mall, Stanford, CA 94305, USA\\
$^{16}$ H.H. Wills Physics Laboratory, Tyndall Avenue, Bristol BS8 1TL, UK\\
  }}

\date{Accepted XXX. Received YYY; in original form ZZZ}

\pubyear{2019}

\begin{document}
\label{firstpage}
\pagerange{\pageref{firstpage}--\pageref{lastpage}}
\maketitle
  
\begin{abstract}
  We explore a disc origin for the highly-blueshifted, variable
  absorption lines seen in the X-ray spectrum of the Narrow Line
  Seyfert 1 galaxy IRAS13224-3809. The blueshift corresponds to a
  velocity of $\sim 0.25$\,c. Such features in other Active Galactic
  Nuclei are often interpreted as UltraFast Outflows (UFOs). The
  velocity is of course present in the orbital motions of the inner
  disk. The absorption lines in IRAS13224-3809 are best seen when the flux
  is low and the reflection component of the disk is strong relative
  to the power-law continuum. The spectra are consistent with a model
  in which the reflection component passes through a thin, highly-ionized absorbing layer
  at the surface of the inner disc, the blue-shifted side of which dominates the flux due to relativistic
  aberration (the disc inclination is about 70 deg). No fast outflow need occur beyond the disc.
\end{abstract}

\begin{keywords}
X-rays: accretion, accretion discs – black hole physics – galaxies: Seyfert
\end{keywords}

\section{Introduction}

Outflows are commonly seen around luminous accreting black holes. The
gravitational energy released by accretion not only leads to radiation
over a wide energy band but also fast winds which, in the case of
supermassive black holes, can drive Active Galactic Nucleus (AGN) Feedback
(for a review see Fabian 2012). The power in an outflow scales as the
wind velocity cubed, $v^3$, so there is considerable interest in
UltraFast Outflows (UFOs) inferred from blueshifted X-ray absorption
features found in a growing number of Seyfert AGN (Pounds et al. 2003,
2016;  Reeves et al. 2009; Tombesi et al. 2010, 2013). The velocities
derived from the absorption features are typically in the range 0.1
-- 0.3 c.

UFO-like features have recently been discovered in the XMM-Newton
(hereafter XMM) EPIC-pn spectrum of the redshift $z=0.0658$ Narrow-Line Seyfert 1 (NLS1)
galaxy IRAS13224-3809 (Parker et al. 17a). An absorption line at about
8.5~keV, plausibly due to FeXXVI or XXV, implies a line of sight
velocity of 0.2--0.25c, respectively.  Blueshifted absorption features
due to oxygen and neon are also seen in the XMM RGS spectrum (Pinto et
al. 2018) and features due to magnesium, silicon and sulphur emerged
from a Principal Component Analysis (PCA: Parker et al. 17b) as well
as direct spectral fitting (Jiang et al. 2018) and the RMS spectrum of the source (Parker et al. 2020).

A remarkable aspect of the X-ray absorption features is that they are
much stronger in the lower half of the large flux range exhibited by
the highly variable source (Parker et al. 17a, Pinto et al. 2018, Jiang
et al. 2018). This could be due to the gas becoming increasingly
photoionized as the source luminosity increases or to geometry if the
brightening is associated with the X-ray emitting corona increasing in
height above the disk (Pinto et al 2018).

A radical alternative interpretation is that the absorbing gas is not
outflowing at all but merely circulating at the surface of the
accretion disc (Gallo \& Fabian 2011 hereafter GF11, 2013). The
orbital velocity of the disk increases up to 0.5\,c at the Innermost
Stable Circular Orbit (ISCO) and so all velocities below that are
present in the disc. Observable UFO velocities of $0.2-0.3$\,c are
reached within radii of $10r_{\rm g}$ around a black hole, where
$r_{\rm g}=GM/c^2$. Doppler beaming makes the approaching side of the
disc brightest, hence absorbing corotating gas along our line of sight
to those radii of the disk can produce a blueshifted absorption line
(GF11).

In this paper we  apply this model to spectra of the lowest and  highest flux states  of  IRAS\,13224-3809.

\section{Simple X-ray spectral modelling of disk absorption in
  IRAS13224-3809}

We modelled the 2016 XMM flux-resolved EPIC-pn spectra of
IRAS~13224$-$3809 between 1 and 10~keV in order to focus on the broad
iron absorption feature seen at 8~keV (Parker et al. 2017a) and a series of
blueshifted absorption features at 1--4~keV (i.e. Si~\textsc{xiv},
S~\textsc{xvi} lines;  Pinto et al. 2018, Jiang et al. 2018). The
data have been gathered into 3 spectra representing low, medium and high
flux, each with a similar number of counts (Parker et al. 2017a).

First, we use the disc emission model from (Jiang et al. 2018). This involves the high density disc reflection model \texttt{xillverd} (Garcia et al. 2016) with the electron density fixed at $n_{\rm e}=10^{19}$\,cm$^{-3}$ during the fit (Jiang et al. 2018). The disc emission consists of two components: one from the inner region with an absorption layer on top (\texttt{xillverd1}) and the other one from the outer region that is not affected by absorption (\texttt{xillverd2}). We assume the same density and ionisation state for the two regions of the disc.

Second, two photoionised-plasma absorption models \texttt{xstar} are constructed and fitted in XSPEC. These absorption models are only applied to \texttt{xillverd1}. The \texttt{xstar} models assume solar abundances except for that of iron and an ionizing luminosity of $10^{43}$\,erg\,s$^{-1}$. A power law with $\Gamma=2$ is assumed for the ionizing spectrum. The redshift parameter ($z$) of the absorption models is fixed at the source redshift ($z=0.066$). Free parameters are the ionization of the plasma ($\xi^{\prime}$\footnote{The prime symbol is to distinguish the parameters of the absorbers from those of the disk reflection models.}), the column density ($N_{\rm H}$), and the iron abundance ($Z_{\rm Fe}^{\prime}$, which is linked to that of the disk). Note that the solar abundances of \texttt{xstar} and \texttt{xillverd} are both taken from (Grevesse \& Sauval 1998).

Third, the \texttt{relconv} model (Dauser 2013) is used to apply relativistic effects to the rest-frame reflection spectra. The total model is \texttt{ tbabs * (relconv*xstar1*xstar2*xillverd1 + relconv*xillverd2 + powerlaw)} in the XSPEC format. The inner radius parameter of the first \texttt{relconv} is set at the radius of the ISCO ($R=1.5$\,$r_{\rm g}$ for a = 0.98) while the outer radius is a free parameter. The inner radius of the second \texttt{relconv} equals the outer radius parameter of the first, and its outer radius is fixed at 400\,$r_{\rm g}$. All the other parameters in the \texttt{relconv} models are linked.

\begin{table}
    \caption{Best-fit parameters for the low flux state and the high flux state spectra. A single emissivity index is used for \texttt{relconv}, i.e. $q=q_1=q_2$. }
    \centering
    \begin{tabular}{ccccc}
      \hline\hline
      Model & Parameter & Unit & Low Flux & High Flux \\
      \hline
      \texttt{relconv} & q & - & >4 & $4\pm2$ \\
                      & $R_{\rm out}$ & $r_{\rm g}$ & <7 & 3 (fixed)\\
                       & $i$ & deg & $71\pm3$ & $70\pm2$ \\
      \texttt{xstar1} & $N_{\rm H,1}$ & $10^{23}$cm$^{-2}$ & $8\pm2$ & $4\pm2$\\
                      & $\log(\xi^{\prime}_{1})$ & log(erg cm s$^{-1}$)& $3.9\pm0.3$ & $3.9\pm0.4$ \\
      \texttt{xstar2} & $N_{\rm H,2}$ & $10^{23}$cm$^{-2}$ & $4^{+2}_{-1}$ & <2 \\
                      & $\log(\xi^{\prime}_{2})$ & log(erg cm s$^{-1}$)& $3.4^{+0.2}_{-0.1}$ & 3.4 (fixed)\\
      \hline
      \texttt{xillverd1}& $\log(\xi)$ & log(erg cm s$^{-1}$)& $1.0\pm0.3$ & $1.0^{+0.1}_{-0.6}$ \\
                        & $Z_{\rm Fe}$ & $Z_{\odot}$ & $3\pm1$ & $4^{+1}_{-2}$\\
                        & norm & - & $5.6^{+0.4}_{-0.8}\times10^{-5}$ & $1.23^{+0.10}_{-0.08}\times10^{-3}$\\
      \hline
      \texttt{xillverd2}& norm & - & <$3\times10^{-6}$ & <$1.5\times10^{-4}$ \\
      \hline
      \texttt{powerlaw} & $\Gamma$ & - & $2.74^{+0.03}_{-0.04}$ & $3.00^{+0.07}_{-0.06}$\\
       & norm & - & $(6\pm1)\times10^{-5}$ & $1.24^{+0.08}_{-0.04}\times10^{-3}$\\
      \hline
      & $\chi^{2}$/$\nu$ & & 112.84/110 & 131.98/120 \\
      \hline\hline
    \end{tabular}
    \label{tab_fit}
\end{table}

\begin{figure*}
    \centering
    \includegraphics[width=\textwidth]{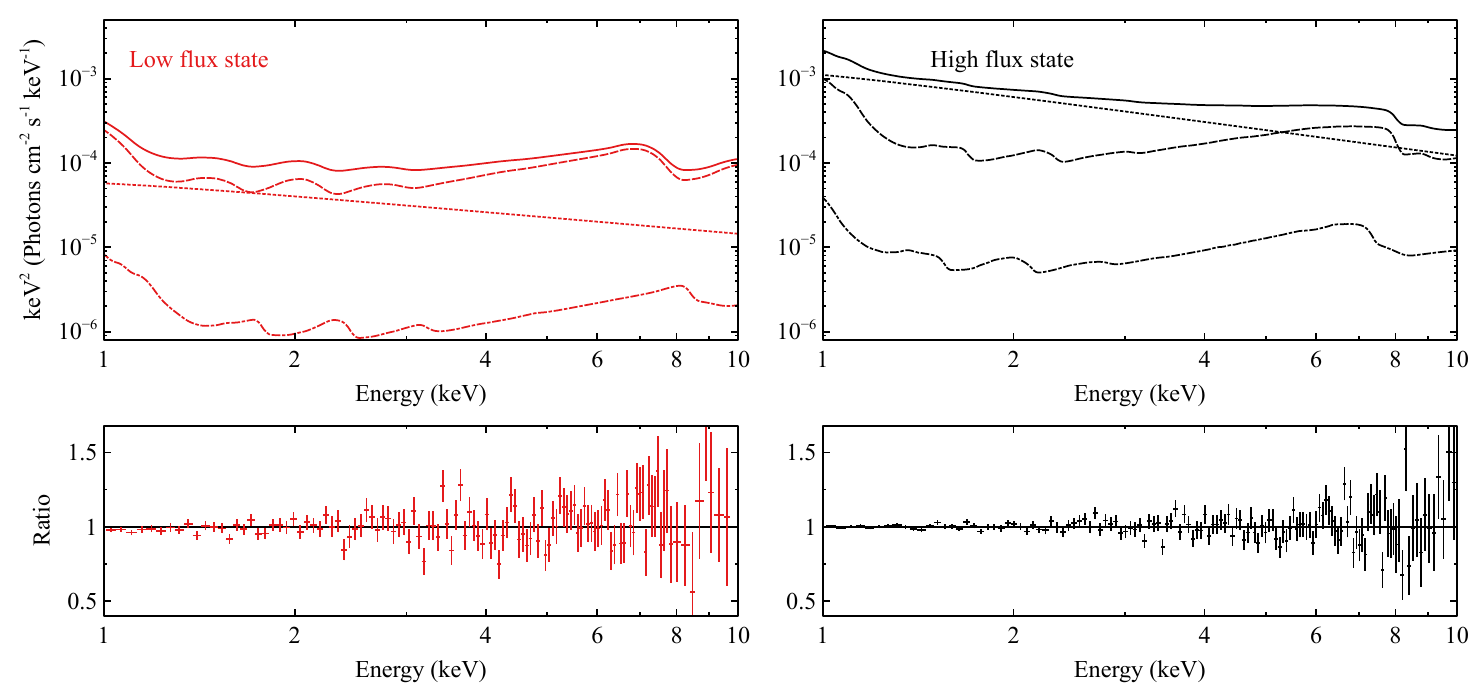}
    \caption{Left top: the best-fit model for the low flux state spectrum (red solid line). Red dashed line: the absorbed reflection emission from the inner region of the disc; red dash-dotted line: the unabsorbed reflection emission from the outer region of the disc; red dotted line: power-law continuum emission. Left bottom: the data/model ratio plot for the low flux state spectrum. Right: same as the left panels but for the high flux state spectrum.}
    \label{fig_fit}
\end{figure*}

\begin{figure*}
    \centering
    \includegraphics[width=\textwidth]{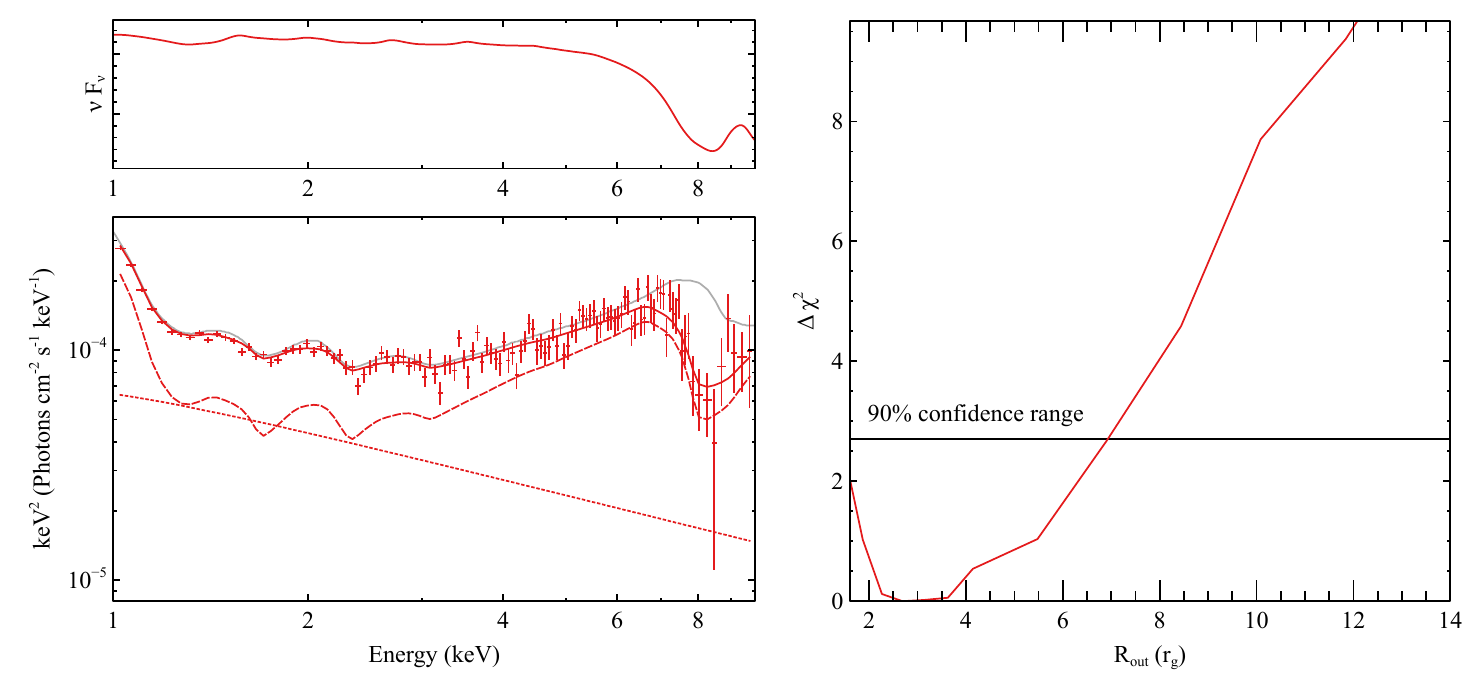}
    \caption{ Top left: the best-fit absorption model for the low flux state spectrum applied to a power law with $\Gamma=2$ to show the shape of the absorption lines. The normalization of the model is arbitrary in this figure. Bottom left: the best-fit model for the low flux state spectrum (red solid line). Only the reflection from the inner disc (dashed line) and the power-law emission (dotted line) are shown for clarity.  The unfolded low flux state spectrum is shown as red crosses for comparison with the model. Right: $\chi^{2}$ as a function of the maximum disk radius of the absorption zone. The inner radius of the annular absorber is fixed at the ISCO (1.5\,$r_{\rm g}$).}
    \label{fig_low_flux_v2}
\end{figure*}

The best-fit models and the corresponding ratio plots can be found in Fig.1, and the best-fit parameters are shown in Table 1. The models describe both the low flux state and the high flux state spectra very well. 

For the low flux state spectrum, the best-fit power-law continuum has a photon index of 2.74. The ionization of the disk reflection component is low with $\xi=10$\,erg\,cm\,s$^{-1}$ and an iron abundance of $\approx3$ compared to solar. A similar high disk inclination (71\,deg) is obtained as for previous work (i.e. $i=67\pm3$\,deg obtained by fitting with \texttt{reflionx}, Jiang et al. 2018). As shown in the top left panel of Fig. 2, the flux of the reflection from the inner region (\texttt{xillverd1}) is more than 10 times higher than the reflection from the outer region (\texttt{xillverd2}). The former is exposed to an ionised absorption layer of $N_{\rm H,1;2}=4-8\times10^{23}$cm$^{-2}$ and $\log\xi=3.4-3.9$ in the rest frame of the disc. The column density of these two absorbers is much higher than the values obtained by the outflow model (e.g. $N_{\rm H}\approx10^{22}$cm$^{-2}$, Jiang et al 2018).

In order to better demonstrate how disk absorption works, we apply our best-fit absorption models (\texttt{kdblur1*xstar1*xstar2}) to a power law with $\Gamma=2$ and show the model in the top panel\footnote{A power law with $\Gamma=2$ would be a horizontal line in the figure. This choice of photon index is only for illustration. We adopt an arbitrary normalization for the power-law model.} of Fig. 2. The best-fit absorption model predicts very broad Fe~\textsc{xxv}/\textsc{xxvi} absorption lines, which correspond to the line features between 8--9\,keV in the real data. The red wing of the broad Fe~\textsc{xxv}/\textsc{xxvi} absorption line extends to 6\,keV, and affects the blue wing of the broad Fe K emission line in the reflection component. In the left bottom panel, we show the best-fit model in comparison with the unfolded low flux state spectrum.
The right panel of Fig. 2 shows $\chi^{2}$ as a function of the maximum disk radius of the absorption zone. The best-fit outer radius of the absorbers is around $3$\,$r_{\rm g}$ with a 90\% confidence range of <7\,$r_{\rm g}$, which suggests the absorbing zone is in the innermost region of the disk within a few $r_{\rm g}$.
 
For the high flux state spectrum, the best-fit power-law continuum has a softer photon index ($\Gamma=3$) compared with the low flux state spectrum. The ionization of the disk reflection component remains consistently low. The relativistic iron line shows an emissivity profile that is consistent with a power law of $q=4$. A lower column density  is required for the first absorber, and a larger uncertainty of the ionization is found. By fixing the ionization of the second absorber at the best-fit value for the low flux state spectrum, we obtain an upper limit of the column density, suggesting weaker absorption in the high flux state. This result is consistent with previous work (Parker et al. 2017a, Pinto et al. 2018, Jiang et al. 2018). The outer radius of the absorption zone is not constrained due to the low reflection fraction in the high flux state (compare the models for two flux states in Fig.\,\ref{fig_fit}), and thus $R_{\rm out}$ is fixed at the best-fit value for the low flux state spectrum.

\section{A physically-motivated  model}

\begin{figure}
\includegraphics[width=\columnwidth]{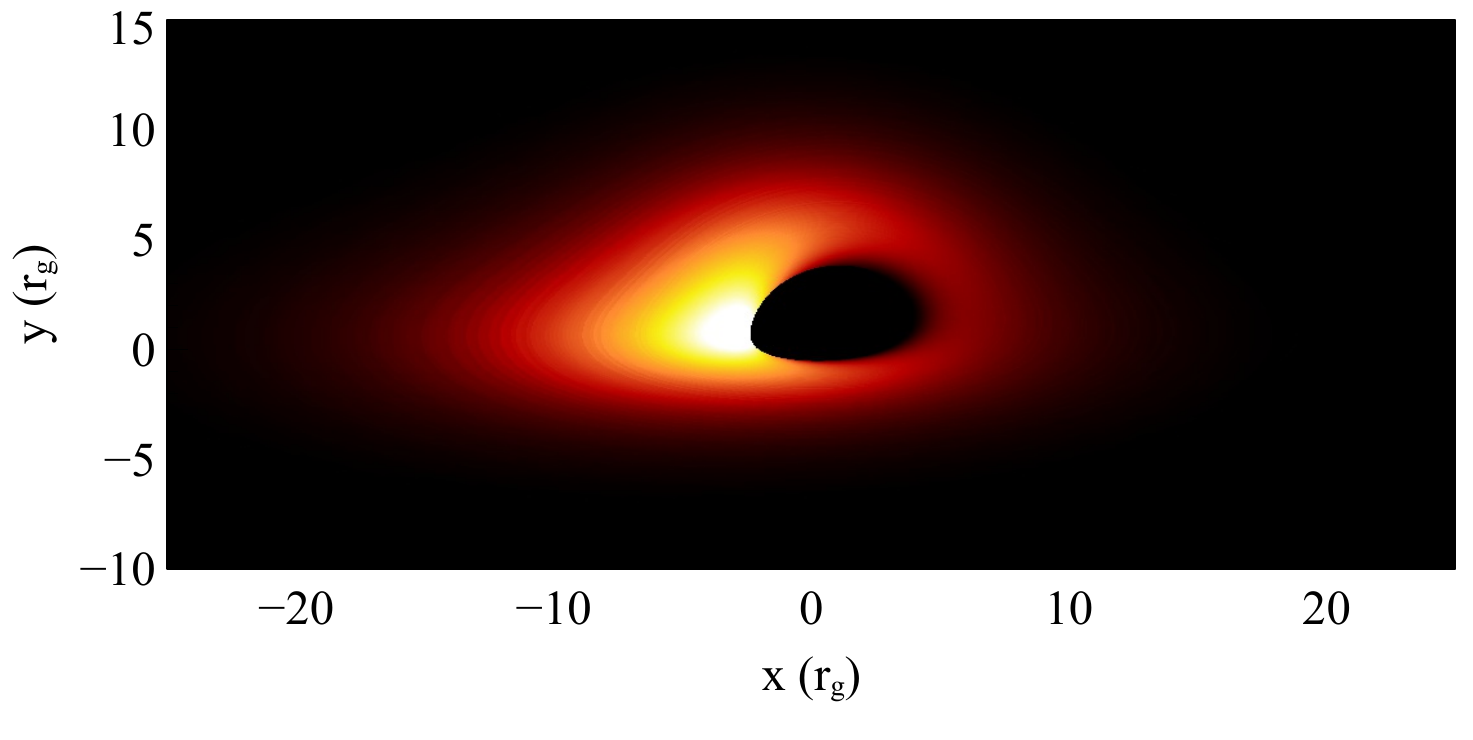}
\includegraphics[width=\columnwidth]{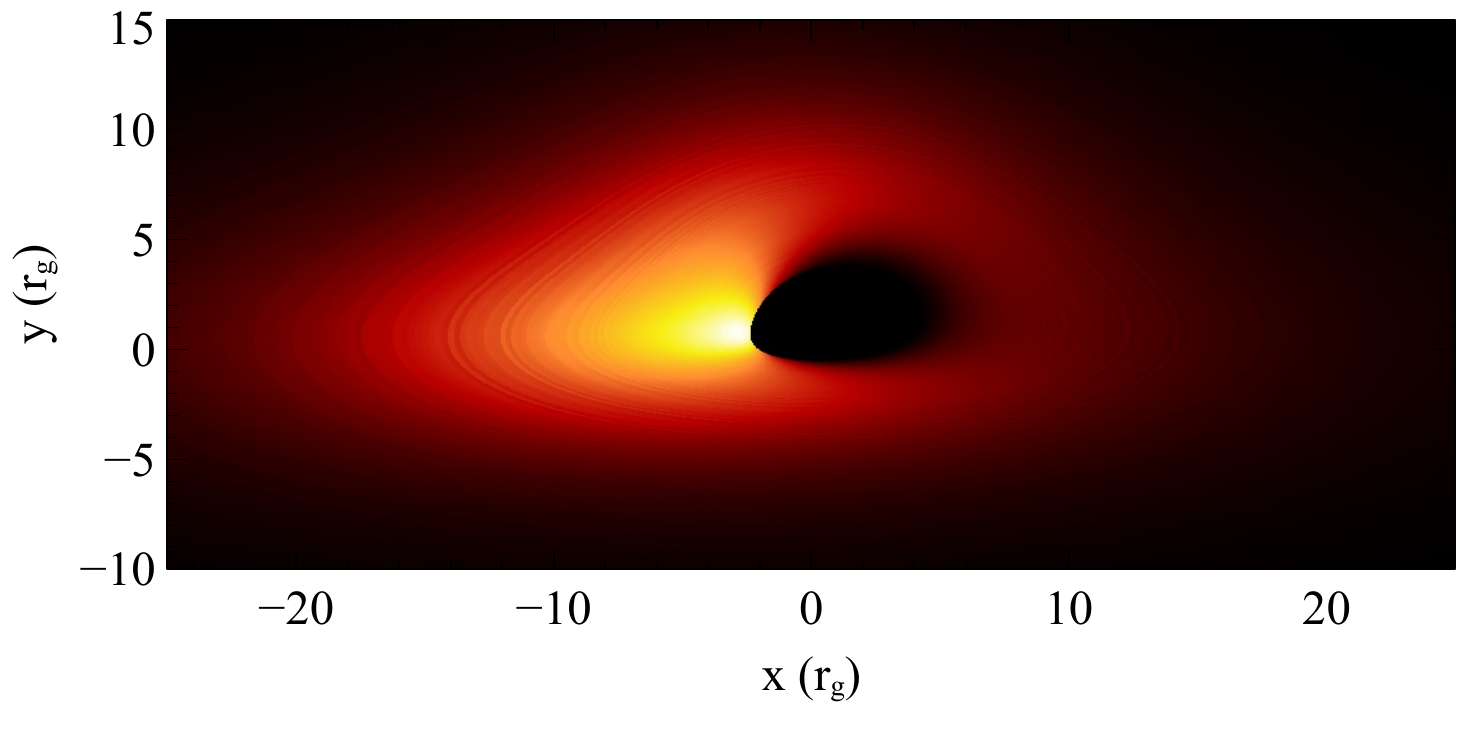}
\includegraphics[width=\columnwidth]{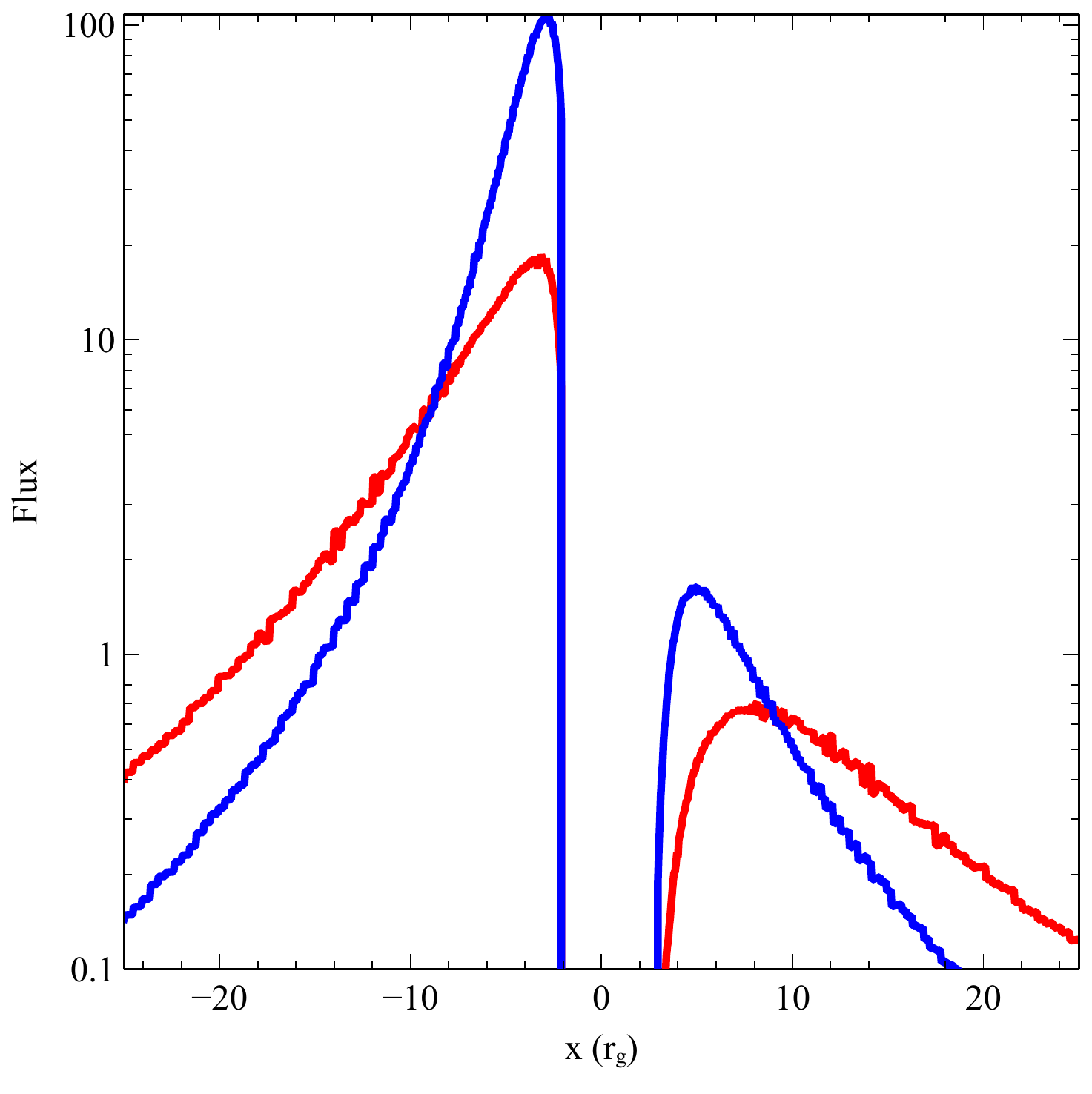}
\caption{Top: Intensity-weighted image of a disc irradiated by a corona
  at height $h= 4r_{\rm g}$, appropriate for the low-flux state. The
  colour table is logarithmic spanning a range of 1000 from white to
  dark red. $x$ and $y$ are in the image plane. Centre: Similar plot for $h= 10r_{\rm g}$ to represent
  the high flux state. Bottom: Flux profiles along a midplane
  cut. Blue line is $h=4 r_{\rm g}$ lamppost; red line is
  $h=10r_{\rm }$.  }
  \label{fig:BHimage}
\end{figure}

\begin{figure}
  \centering{
\includegraphics[width=\columnwidth]{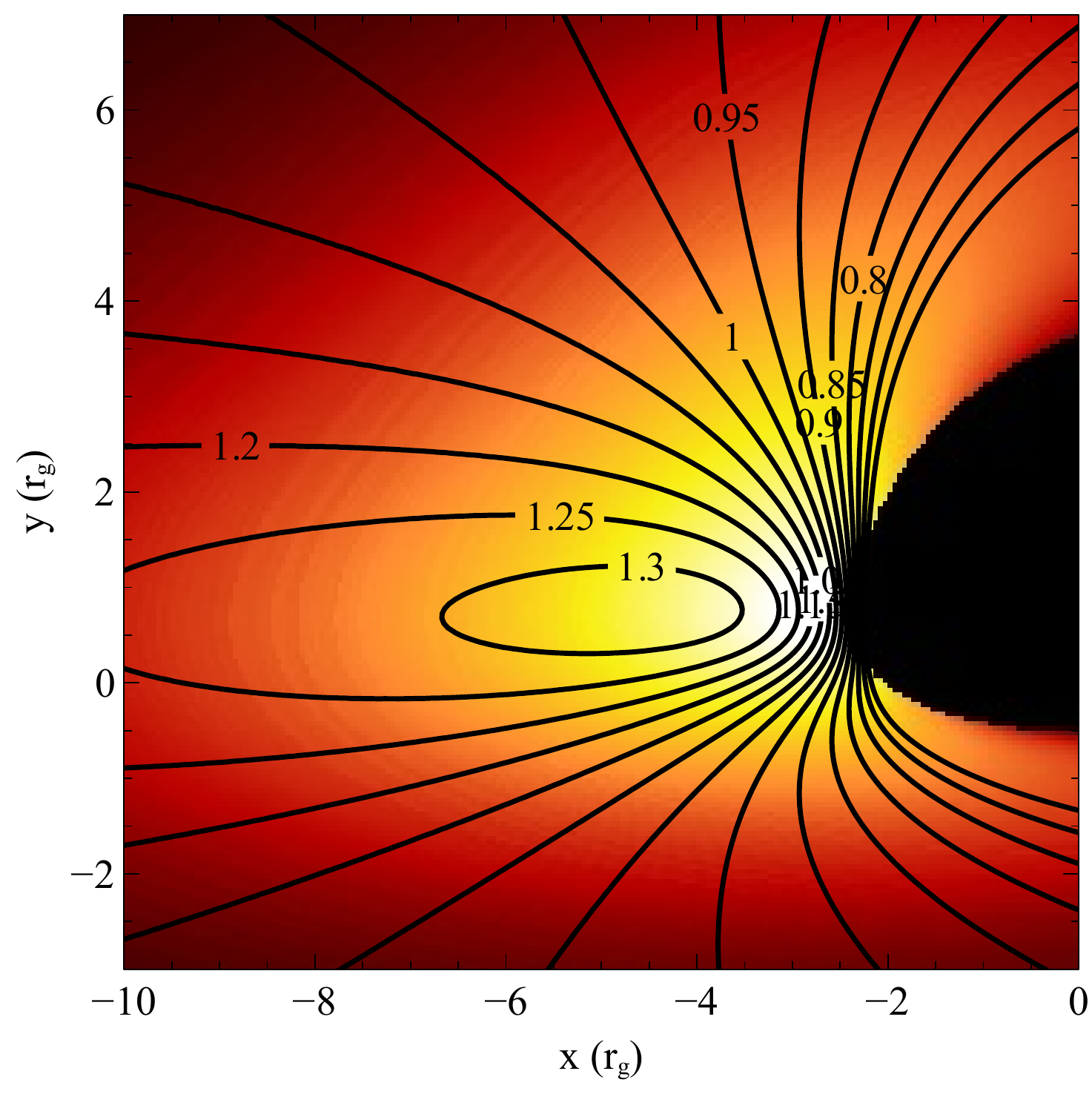}}
\caption{ The approaching side of the disc shown in the top panel of Fig.~3 overlaid with contours showing the g-factor at the surface of the disc in the rest-frame of the source. $g=E_{\rm obs}/E_{\rm em}$ and is about 1.27 at the line centre. E is photon energy, $x$ and $y$ are in the image plane.}
\label{fig:gcon}
\end{figure}

\begin{figure}
  \centering{
\includegraphics[width=\columnwidth]{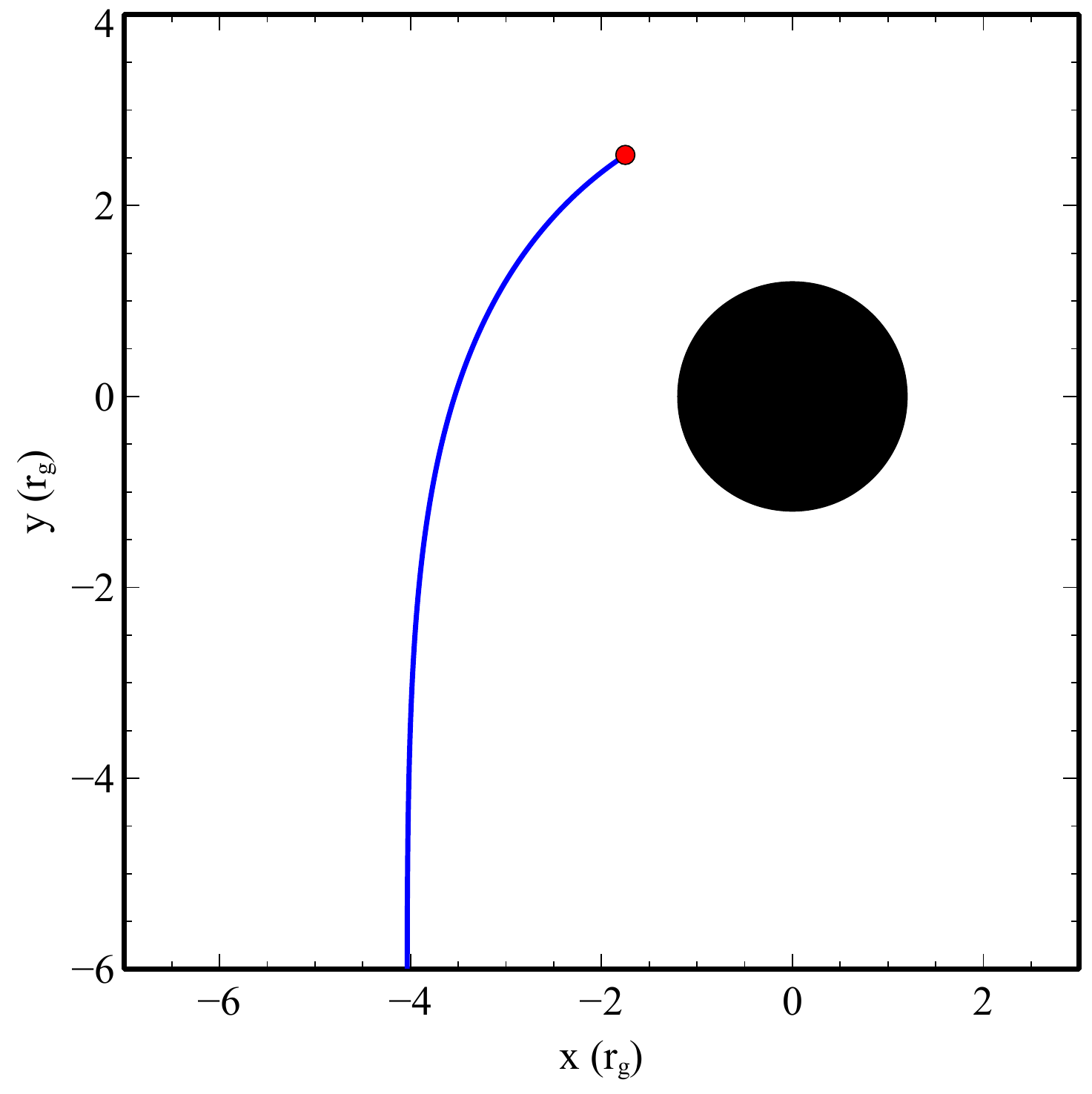}}
\caption{Path of a photon from the brightest part of the reflection
  emission region (Fig. 3 top). The observer lies off the bottom. $x$ and $y$ here are in the frame of the disc.}
  \label{fig:atmos}
\end{figure}

The above modelling  shows that  highly-ionized absorption  applied
only to the reflection component is consistent with the  low-flux spectrum. Such absorption becomes
undetectable in the high state  due to a combination of a) 
reduction in the column density of the absorber and b) dilution by the
power-law continuum, which becomes relatively stronger.  The 1--10~keV
power-law continuum increases by a factor of 13 from low to
high-state, whereas the reflection component increases by a factor of
only 3. At 8.5 keV where the Fe-K absorption feature appears, the
factors are 6.9 and 2.1, respectively (see Fig.~1).

We assume that the 
absorbing gas shares the orbital velocity of the disc immediately
below it. Fig.~2 (right) shows that the detectable absorber extends in radius from about 2 to $7r_{\rm g}$. (There may be undetected absorption at larger radii from lower column-density material.) An image (computed using the code from Reynolds et
al. 1999) showing how the region appears to the observer is shown in
Fig. 3: the disk is irradiated by a corona (not shown) located along
the spin axis at a height of $\sim5r_{\rm g}$ above its centre (Alston et al. 2020).  The
brightest part of the reflection corresponds to the region with the
highest blueshift which also lies between about 2 and $8r_{\rm g}$ from
the black hole along the {\em apparent} approaching midline of the
disc (Figs.~3 and 4). It is {\em apparent} because the path of a photon appearing to the observer to originate from the brightest part of the disc originates behind the black hole, due to light bending in the Kerr metric (see Fig.~5).   

The high flux state corresponds to the coronal height being about
$10r_{\rm g}$. Its radius will also be larger. The reflection component
then no longer originates from a small part of the inner disc but is
more spread out (Fig.~3 middle and lower plots), contributing further to the lack of an observed absorption feature in the high state.

A high inclination is required  for UFO-like
features to be seen. At the inclination of $\sim$\,71 deg inferred for
IRAS13224-3809, the total {\it vertical} column density of absorbing matter in the low state is $\sim 4\times 10^{23}$\,cm$^{-2}$, which corresponds to a Thomson depth\footnote{\texttt{xstar} does not currently include the effects of electron scattering, which will slightly affect the inferred column densities.} of about 0.3.  It may be less in the
high state. (Any contribution by the absorber to the total reflection
component can therefore be neglected.) The absorber has an ionization
parameter about 1000 times higher than the disc, so the absorption
density must be about $10^{16}$cm$^{-3}$. The depth of the absorbing region is then only $\sim3\times 10^7$\,cm, or $\sim10^{-4}r_{\rm g}$, if the black hole mass is $2\times10^6\Msun$ (Alston et al 2020).  Our fits require 2 absorbers representing a factor of 3 range of ionization parameter which  means a corresponding range of density in the absorbing layer. We assume that the absorbing layer lies immediately above the surface of the dense disc.

Absorption is imprinted on the reflection spectrum both as radiation enters the disc and as it leaves it. Assuming that much of the incoming radiation is incident vertically on the disc and that the observed radiation is from an angle near 70 deg, then 3/4 of the total absorbing column density traversed is found as it leaves the disc. This means that the above depth is overestimated by about 30 per cent. 

\section{Discussion}

We conclude that a plausible origin for the variable blueshifted
absorption in IRAS13224-3809 lies in a thin, low density, outer layer of 
 the inner accretion disc. Our line of sight to the brightest part
of the reflection component passes through regions above the disc
where absorbing matter corotating with the disc would appear blueshifted by the observed value.

The model may apply to other detected UFOs, particularly if strong,
compact, inner reflection components are present and the disk
inclination and spin are high.  UFOs reported from objects which have
no reflection component are likely to be genuine outflows, such as the
case for the highly-blueshifted absorption seen in UltraLuminous X-ray
sources (ULX: Pinto et al. 2016, Walton et al. 2017, Kosec et al.
2018). Moreover, any low density absorbing gas must not be so
highly-ionized that few ions are available to cause absorption
features.  This may occur in luminous stellar mass black hole systems
where the disc tends already to have a high ionisation parameter and any lower density gas above it would be even more highly ionised.

The situation for observing ultrafast disc absorption features is optimised
when the brightest part of the reflection component is highly
localised on the approaching side of the disk and the inclination is
high enough for our line of sight to intercept a sufficient column density of absorber above the disc.  This implies a compact corona at a small height above the black hole.

Although IRAS13224-3809 may be above the Eddington limit, much of the radiation in the inner region originates in the corona, which lies above  the disc and radiates down on it. This geometry minimises radiation-driven outflows from the disc. 

The RMS spectrum of IRAS\,13224-3809 (Parker et al. 2020) shows that the absorption line is strongly variable, 30 per cent more than the neighbouring continuum. This may reflect rapid variations in the disc surface.   

Disc absorption presents a new feature of X-ray reflection. The
highly blueshifted features found in IRAS13224-3809 represent a probe
of the innermost accretion flow {\em within a few gravitational radii of the black hole}.  High resolution spectra obtained by future missions,
such as XRISM and Athena, will potentially resolve the line shape of
the broad absorption line features, locating and mapping the disc absorber with high accuracy.

\section*{Acknowledgements}
We thank the referee for helpful comments.
WNA and ACF  acknowledge support from ERC Advanced Grant FEEDBACK,
340442. JJ acknowledges support from the Tsinghua Astrophysics Outstanding Fellowship and the Tsinghua Shuimu Scholar Programme. JJ also acknowledges support from the Cambridge Trust and the
Chinese Scholarship Council Joint Scholarship Programme
(201604100032). CSR thanks the UK Science and Technology Facilities Council (STFC) for support under the New Applicant grant ST/R000867/1, and the European Research Council (ERC) for support under the European Union's Horizon 2020 research and innovation programme (grant 834203). PK acknowledges support from the STFC. BDM acknowledges support from the European Union’s Horizon 2020 Research and Innovation Programme under the Marie Skłodowska-Curie grant agreement No. 798726. GM acknowledges funding by the Spanish State Research Agency (AEI)
Project No. ESP2017-86582-C4-1-R and No. MDM-2017-0737 Unidad de
Excelencia Maria de Maeztu - Centro de Astrobiologica (CSIC-INTA).

%\bibliographystyle{mnras}
%\bibliography{refs}

% Don't change these lines
\bsp	% typesetting comment
\label{lastpage}
\end{document}